\documentclass[%
aps,
reprint,
superscriptaddress,
amsmath,amssymb,
prl,
floatfix,
showkeys
]{revtex4-2}

\usepackage[utf8]{inputenc} 
\usepackage[T1]{fontenc}
\usepackage{microtype}

\setcitestyle{numbers,square,citesep={,\kern-.24em}}

\usepackage{graphicx}
\usepackage{dcolumn}
\usepackage{bm}
\usepackage{float}
\usepackage[colorlinks,urlcolor=Blue,linkcolor=Blue,citecolor=Blue]{hyperref}
\usepackage{gensymb}
\usepackage{textcomp}

\mathchardef\mhyphen="2D

\usepackage{CJK}
\usepackage{upgreek}

\usepackage[dvipsnames]{xcolor}

\usepackage[
top=0.70in, left=0.75in, right=0.75in, bottom=0.71in,
]{geometry}

\newcommand{\ket}[1]{\ensuremath{\left|#1\right>}}

\newcommand{\op}[1]{\hat{#1}}

\def\be{\begin{equation}}
\def\ee{\end{equation}}

\begin{document}

\title{Atom Interferometry with Floquet Atom Optics}

\begin{CJK*}{UTF8}{gbsn}

\author{Thomas Wilkason}
\thanks{These authors contributed equally to this work.}
\affiliation{
  Department of Physics, Stanford University, Stanford, California 94305, USA
}

\author{Megan Nantel}
\thanks{These authors contributed equally to this work.}
\affiliation{
Department of Applied Physics, Stanford University, Stanford, California 94305, USA
}%

\author{Jan Rudolph}
\thanks{These authors contributed equally to this work.}

\affiliation{
  Department of Physics, Stanford University, Stanford, California 94305, USA
}%

\author{Yijun Jiang (姜一君)}
\affiliation{
Department of Applied Physics, Stanford University, Stanford, California 94305, USA
}%

\author{\mbox{Benjamin E.\ Garber}}
\affiliation{
  Department of Physics, Stanford University, Stanford, California 94305, USA
}%

\author{Hunter Swan}
\affiliation{
  Department of Physics, Stanford University, Stanford, California 94305, USA
}%

\author{Samuel P.\ Carman}
\affiliation{
  Department of Physics, Stanford University, Stanford, California 94305, USA
}%

\author{Mahiro Abe}
\affiliation{
  Department of Physics, Stanford University, Stanford, California 94305, USA
}%

\author{Jason M.\ Hogan}
\email[]{hogan@stanford.edu}
\affiliation{
  Department of Physics, Stanford University, Stanford, California 94305, USA
}%

\date{24 October 2022}
\microtypesetup{disable}
\begin{abstract}
Floquet engineering offers a compelling approach for designing the time evolution of periodically driven systems. We implement a periodic atom-light coupling to realize Floquet atom optics on the strontium ${}^1\!S_0\,\text{--}\, {}^3\!P_1$ transition. These atom optics reach pulse efficiencies above $99.4\%$ over a wide range of frequency offsets between light and atomic resonance, even under strong driving where this detuning is on the order of the Rabi frequency. Moreover, we use Floquet atom optics to compensate for differential Doppler shifts in large momentum transfer atom interferometers and achieve state-of-the-art momentum separation in excess of $400~\hbar k$. This technique can be applied to any two-level system at arbitrary coupling strength, with broad application in coherent quantum control.
\end{abstract}

\maketitle

\end{CJK*}
\microtypesetup{enable}
Periodic driving has been employed for coherent control of quantum systems ranging from single atoms to many-body and solid-state systems, and can reveal new nonequilibrium phenomena~\cite{Bukov2015, Eckardt2017,Oka2019,Weitenberg2021}. In addition to facilitating the spectral design of dressed states, Floquet modulation can be used to engineer the time-domain dynamics of two-level systems in the presence of a strong drive~\cite{Deng2015,Deng2016}. This finds application in quantum information, where periodic driving can increase the robustness of quantum superpositions against decoherence and dephasing~\cite{Huang2021, Gandon2021}, as well as improve the fidelity of quantum gate operations~\cite{Shi2016,Wu2021}. Such high-fidelity state manipulations are closely related to atom optics used in light-pulse atom interferometry, where an atomic system is interrogated with light at optical frequencies to introduce inertial sensitivity~\cite{Kasevich1991,Giltner1995}. Highly efficient atom optics allow for the application of many sequential light pulses, enabling large momentum transfer (LMT) atom interferometers with increased space-time area and sensitivity~\cite{McGuirk2000,Muller2008,Chiow2011,Plotkin-Swing2018}.

In an LMT interferometer, the atomic center-of-mass wave function is split into two components that follow distinct trajectories, with the separation between these arms determined by the number of photon momenta transferred by the atom optics.  For fast atom optics pulses, the bandwidth can be sufficiently broad to efficiently drive transitions on both interferometer arms, despite the differential Doppler shift~\cite{Rudolph2020}. However, continuing to scale up the momentum separation leads to a substantial loss in transfer efficiency when the Doppler shift $\Delta$ approaches the atom-light coupling strength $\Omega$. There are a number of well-established strategies to address such frequency errors in two-level systems, including composite pulse sequences~\cite{Butts2013, Dunning2014, Berg2015}, adiabatic rapid passage (ARP) techniques~\cite{Marte1991,Bateman2007, Kotru2015}, and optimal quantum control protocols~\cite{Saywell2018,Saywell2020}. In many cases, it is of interest to design pulses that provide broadband excitation to accommodate a continuum of detuning errors, such as Doppler shifts due to the temperature of an ensemble of atoms. Instead, we study Floquet-engineered atom optics designed to achieve perfect transfer efficiency for a discrete set of detuning errors, corresponding to the two velocities of the atom interferometer arms.  We show that Floquet atom optics can compensate for a wide range of Doppler detunings in both the strong ($\Omega\sim|\Delta|$) and weak ($\Omega\ll|\Delta|$) coupling regimes, thereby circumventing a leading loss mechanism in LMT atom interferometry.

We consider a periodic Hamiltonian using light on resonance with an optical transition in a two-level system at rest in the lab frame. Using the rotating wave approximation (RWA) to neglect counterrotating terms at optical frequencies, the Hamiltonian describing one of the interferometer arms with Doppler detuning $\Delta$ is
\begin{align*}
   \hat{H}(t) = \frac{\hbar\,\Delta}{2}\hat{\sigma}_z + \frac{\hbar\,\Omega(t)}{2} \hat{\sigma}_x,
\end{align*}
where $\Omega(t)$ is some real, time-dependent Rabi coupling, and $\op{\sigma}_j$ are the Pauli operators~\footnote{Here we use the convention where $\op{\sigma}_z\ket{g} = \ket{g}$ and $\op{\sigma}_z\ket{e} = -\ket{e}$}.  We periodically modulate the amplitude of the light pulse envelope at some period $T$, such that $\Omega(t)=\Omega(t+T)$.  The fundamental modulation frequency $\beta \equiv 2\pi/T$ is generally detuned by some modulation offset $\delta \equiv \beta - |\Delta|$ with respect to the Doppler-shifted resonance. Note that despite making the RWA at optical frequencies, counter-rotating terms at harmonics of $\beta$ cannot be ignored in the strong coupling regime ($\Omega \sim |\Delta|$).  As a result, we effectively realize the dynamics of the strong drive Rabi model~\cite{Fuchs_2009, Dai2017}.

\begin{figure*}[t]
         \includegraphics[width=\linewidth]{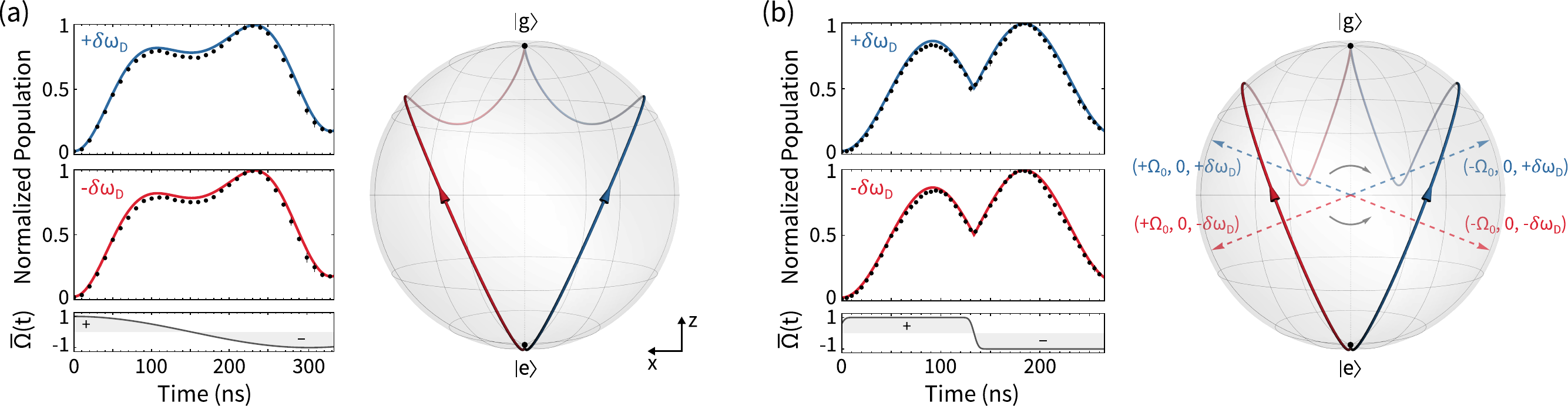}
     \caption{(a) Measured normalized ground state population for a sine-wave Floquet pulse (black points) and associated pulse envelope $\overline{\Omega}(t) \equiv \Omega(t)/\Omega_0$ (bottom panel) with modulation offset $\delta = -2\pi \times 0.35~\text{MHz}$, detuning $\Delta = \pm \delta\omega_D= \pm 2\pi\! \times \! 2~\text{MHz}$, and Rabi frequency $\Omega_0 = 2\pi \times 5~\text{MHz}$.  Numerical density matrix solutions are shown for both positive (blue) and negative (red) detuning with no free parameters.  These state trajectories are also illustrated on the Bloch sphere for an ideal Floquet $\pi$ pulse without spontaneous emission. (b) Square-wave Floquet pulse with $\delta = 2\pi \times 1.75~\text{MHz}$, $\Delta = \pm 2\pi\! \times \! 2~\text{MHz}$, and $\Omega_0 = 2\pi \times 5~\text{MHz}$. The blue and red dashed arrows show the torque vectors for the associated Doppler detunings, which flip from left to right at $t=T/2$ as suggested by the gray arrows.
     }
    \label{fig:1}
\end{figure*}

In our case, an interferometer arm has a lab-frame Doppler shift of either $\Delta = +\delta\omega_D$ or $\Delta = -\delta\omega_D$. For any purely amplitude-modulated (real) function $\Omega(t)$, the evolution of the two arms on the Bloch sphere is instantaneously mirror symmetric~\footnote{That is, if $\{g(t),e(t)\}$ is a solution for $\Delta = +\delta\omega_D$, with $g(t)$ and $e(t)$ the amplitudes of the ground and excited states, respectively, then $\{g(t)^*,-e(t)^*\}$ is a solution for $\Delta = -\delta\omega_D$.}. This symmetry guarantees the two interferometer arms will have the same transfer efficiency for any pulse shape and duration.

Using Floquet's theorem, we can write the time evolution of the state in terms of the two Floquet modes~\cite{Shirley1965}. In the strong coupling regime ($\Omega \sim |\Delta|$), these modes are the time-dependent analogs of the dressed states~\cite{Guerin1997}.  By analogy with Rabi oscillations in the weak coupling limit, we may describe the strong coupling Floquet dynamics as resulting from a diabatic projection of the initial quantum state into a superposition of time-dependent dressed states (the Floquet modes). 

Assuming the system starts in one of the bare atom states $\ket{g}$ or $\ket{e}$, we engineer efficient transitions to the opposite state by analyzing the evolution of the Floquet modes as a function of the modulation offset $\delta$ and pulse duration $t_{\pi}$.  For weak coupling ($\Omega \ll |\Delta|$) and sinusoidal drive $\Omega(t) = \Omega_0 \cos{\beta t}$, we recover the conventional RWA $\pi$-pulse condition $t_{\pi} = \pi/\Omega_g$ with generalized Rabi frequency $\Omega_g = \sqrt{\delta^2 +(\Omega_0/2)^2}$, where maximal efficiency is reached when the corotating frequency component of $\Omega(t)$ is on resonance ($\delta = 0$). In contrast, when $\Omega \sim |\Delta|$ the analogous resonance condition becomes time dependent, reflecting the fact that the Floquet modes periodically exchange energy with the drive.  Nevertheless, for any $\Omega$ and $\Delta$, a simultaneous solution can be found for $\delta$ and $t_\pi$ that satisfies a generalized $\pi$-pulse condition to allow perfect transfer~\cite{Supplemental}. We take advantage of this to engineer optimized Floquet $\pi$ pulses.

We experimentally demonstrate Floquet atom optics on the $689~\text{nm}$ transition in ${}^{88}\text{Sr}$, with $\ket{g}=\ket{{}^1\!S_0}$ and $\ket{e}=\ket{{}^3\!P_1, m = 0}$. First, $10^6$ atoms are prepared at a temperature of $2~\mu\text{K}$ and a cloud size of $\sigma = 135~\mu\text{m}$. The atoms are interrogated with a sequence of alternating laser pulses from opposite directions as in~\cite{Rudolph2020}, each derived from a Ti:sapphire laser that is frequency stabilized to a reference cavity. The pulse envelopes are controlled by two single-pass acousto-optic modulators (AOMs) driven by independent channels of an arbitrary function generator (AFG). The light is delivered to the atoms via optical fibers, and each beam has an optical power of $180~\text{mW}$ with a $1/e^2$ radial waist of $2~\text{mm}$. On resonance, we achieve a Rabi frequency of $\Omega_0 = 2\pi \times 5~\text{MHz}$ and a $\pi$-pulse duration of $t_0 = 100~\text{ns}$. For amplitude-modulated Floquet pulses, the carrier waveform is modulated by a chosen envelope function $\Omega(t)$ using the AFG. To ensure pure amplitude modulation (AM), we interferometrically measure any residual self-phase modulation in the optical fibers and apply a compensatory phase to the waveform~\cite{Supplemental}. To simulate an arbitrary Doppler shift $\Delta=\pm \delta \omega_D$ for atoms at rest in the lab frame, we detune the laser carrier frequency from resonance using the AOMs.

We characterize the dynamics of the two-level system during a Floquet pulse with the following experimental sequence. First, a bias magnetic field of $100~\text{G}$ is applied to suppress unwanted excitations to the $m = \pm 1$ Zeeman sublevels of $\ket{{}^3\!P_1}$. We then transfer the atoms to the excited state $\ket{e}$ using a resonant $\pi$ pulse. Next, we apply a Floquet pulse of variable duration. Before detection, a $500~\text{ns}$ push pulse on the ${}^1\!S_0\,\text{--}\, {}^1\!P_1$ transition at $461~\text{nm}$ leads to vertical separation of the states after time of flight. Finally, we measure the state populations via fluorescence imaging on this same transition.  Our choice to initialize the atoms in $\ket{e}$ when characterizing Floquet $\pi$ pulses minimizes detection errors caused by spontaneous decay during the push pulse.

Figure~\hyperref[fig:1]{1(a)} shows the normalized ground state population during a Floquet pulse with sine-wave modulation $\Omega(t) = \Omega_0 \,\cos{\beta t}$ at laser detunings $\Delta = \pm 2\pi\! \times \! 2~\text{MHz}$. The observed dynamics notably differ from a resonant Rabi oscillation, which is a manifestation of the time dependence of the Floquet modes. We find efficient population inversion $>\!99\%$ at the pulse parameters $t_{\pi} = 232~\text{ns}$ and $\beta = 2\pi \times 1.65~\text{MHz}$. As expected, we observe symmetric time evolution for equal and opposite detunings. The calculated state evolution during the Floquet $\pi$ pulse is shown on the Bloch sphere, illustrating the mirror symmetry for the two cases. We repeat the same characterization for square-wave modulation $\Omega(t) = \Omega_0 \,\text{sgn}(\cos{\beta t})$, where the sign of the pulse envelope is switched every half-period $T/2$ [see Fig.~\hyperref[fig:1]{1(b)}]. Efficient population inversion $>\!99\%$ is achieved at $t_{\pi} = 183.5~\text{ns}$ and $\beta = 2\pi \times 3.75~\text{MHz}$, and we again observe symmetric time evolution. While both approaches achieve comparable fidelity, the optimal pulse duration for square-wave modulation is shorter due to higher rms pulse amplitude~\cite{Supplemental}.

The short duration ($t_\pi < T$) of the pulses used in Fig.~\hyperref[fig:1]{1} suggests a complementary time-domain description of the Floquet dynamics analogous to a composite pulse sequence.  In particular, the piecewise constant amplitude of the square-wave excitation allows for a simple geometric solution of the state evolution on the Bloch sphere. During each half-period of modulation $T/2$, the state vector precesses at a constant rate $\sqrt{\Omega_0^2+\Delta^2}$, tracing out a circular arc on the sphere. This arc is coincident with the base of a cone that has its vertex at the origin and an axis given by the torque vector, $(\Omega_0, 0, \Delta)$ for the first half-period.  Every subsequent half-period, $\Omega(t)$ changes sign, mirroring the torque vector about the $yz$ plane and leading to further precession around the base of the associated cone. Thus, finding optimal pulse parameters is reduced to a purely geometric problem of determining the intersections of a series of cones that pass through the two poles of the Bloch sphere.  For $0 \leq |\Delta| \leq \Omega_0$ a solution exists using two cones:
\begin{align*}
    t_{\pi} &= \frac{2\pi}{\sqrt{\Omega_0^2+\Delta^2}}, &
    \beta &= \frac{\sqrt{\Omega_0^2 + \Delta^2}}{ 1 + \dfrac{2}{\pi}\cos^{-1}\!{\left(\dfrac{\sqrt{ \Omega_0^2 + \Delta^2}}{\sqrt{2}\,\Omega_0}\right)}}
    \label{eq:squarepulse_exp}
\end{align*}
This corresponds to the dynamics shown in Fig.~\hyperref[fig:1]{1(b)}.  For larger detunings, additional half-periods are required to achieve efficient state transfer.  Specifically, for detunings in the range $\cot\!{\left[\pi/2(n-1)\right]} \leq (|\Delta|/\Omega_0) \leq \cot{\left(\pi/2n\right)}$ with integer $n$, a solution can be found with $n$ cones (half-periods).  We find analytic solutions for $t_{\pi}$ and $\beta$ up to $n=5$~\cite{Supplemental}.

 \begin{figure}
           \includegraphics[width=\linewidth]{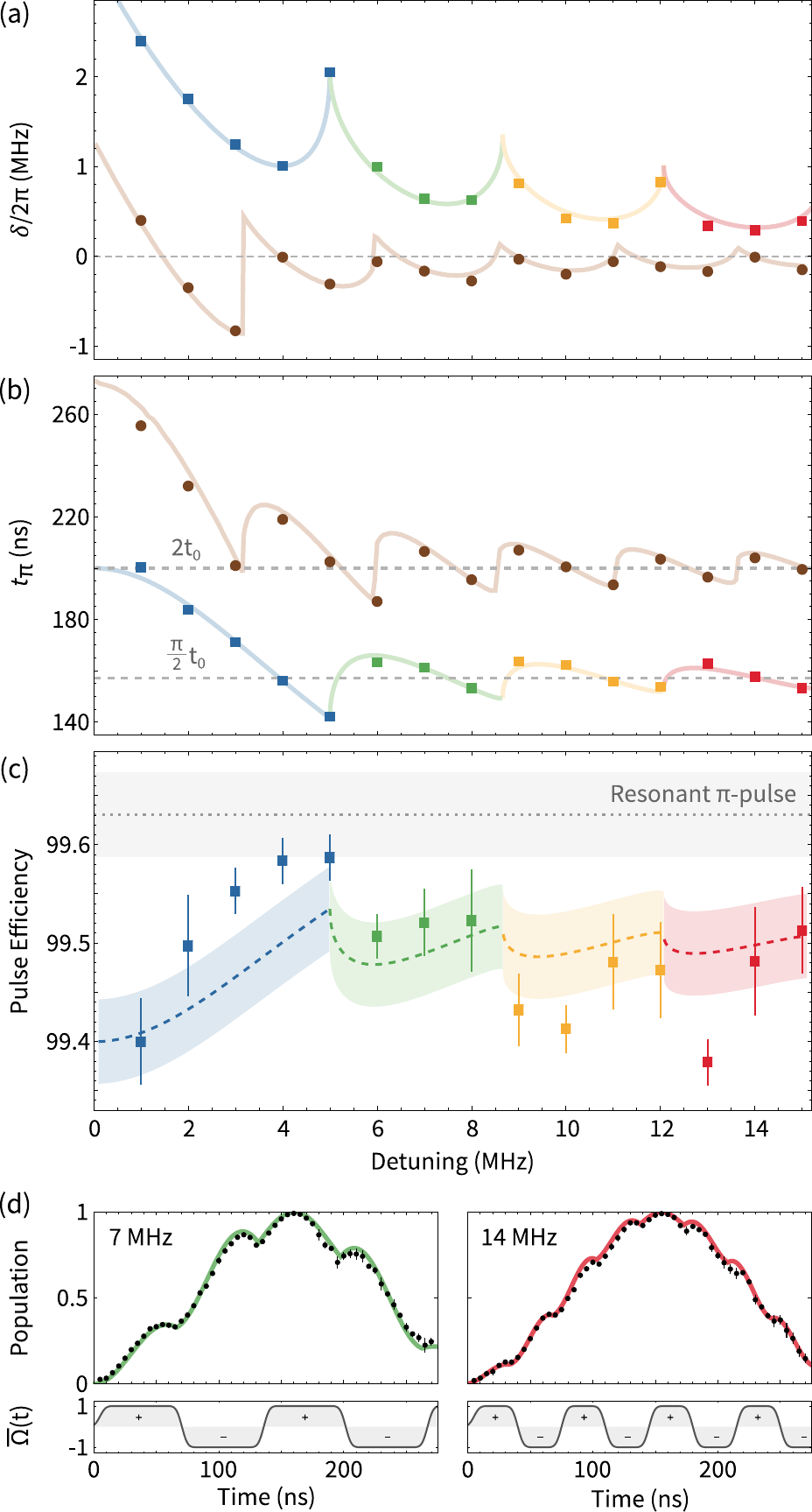}
     \caption{(a) Experimental characterization of optimal Floquet $\pi$-pulse modulation offset $\delta = \beta - |\Delta|$ as a function of laser detuning $\Delta/2\pi$, for sine-wave (circles) and square-wave (squares) AM. (b) Experimentally optimized pulse duration $t_{\pi}$. The colored curves (blue, green, yellow, and red) are the analytic solutions for square-wave modulation, with the color corresponding to the number of half-periods $n$ required to reach optimal transfer ($n = 2$, 3, 4, and 5, respectively). The brown theory curves are numerical solutions of the Schr\"{o}dinger equation for sine-wave modulation. (c) Measured efficiency of square-wave modulated pulses. The dashed, colored curves are the expected pulse efficiency based on the pulse duration, extrapolated from the measured efficiency of resonant $\pi$ pulses (dotted gray line). The error bars and bands show one standard deviation of measurement uncertainty. (d) Example time evolution for $\Delta = 2\pi \times 7~\text{MHz}$ (left) and $2\pi \times 14~\text{MHz}$ (right) with associated numerical Schr\"{o}dinger fits corresponding to $n = 3$ (green) and $n=5$ (red). As the detuning increases, the dynamics converge toward Rabi oscillations at reduced frequency $\frac{2}{\pi} \Omega_0$.}
     \label{fig:2}
 \end{figure}

Figure~\hyperref[fig:2]{2} shows optimal Floquet atom optics parameters as a function of detuning $\Delta$ for both sine-wave and square-wave AM. We determine the maximum measured transfer efficiency with a two-parameter grid search over $\delta$ and $t_\pi$. The cusps in Figs.~\hyperref[fig:2]{2(a)} and \hyperref[fig:2]{2(b)} mark the boundaries of detuning ranges where $n$ half-periods are required for optimal transfer. In the large detuning limit $|\Delta| \gg \Omega_0$, the optimal modulation frequency converges to $\delta = 0$ for both modulation types, as expected in the weak coupling regime where the RWA is valid [see Fig.~\hyperref[fig:2]{2(a)}]. In this regime, the pulse duration converges to the expected $t_{\pi} = 2t_{0}$ for sine-wave modulation and $t_{\pi} = \frac{\pi }{2}t_{0}$ for square-wave modulation [Fig.~\hyperref[fig:2]{2(b)}], corresponding to the Rabi couplings of the respective Fourier amplitudes at the corotating frequency $\beta$~\cite{Supplemental}. We see this same asymptotic behavior in the time domain in Fig.~\hyperref[fig:2]{2(d)}, where for larger detunings the Floquet dynamics begin to resemble Rabi oscillations that are only weakly perturbed by counterrotating terms. Thus, we observe agreement with theory for Floquet $\pi$ pulses across the strong and weak coupling regimes.

Figure~\hyperref[fig:2]{2(c)} shows the measured efficiency of the square-wave Floquet pulses. The dominant source of inefficiency is spontaneous decay during the pulse, and as a result, the square-wave pulses are uniformly more efficient than the best measured efficiencies of the sine-wave pulses due to their shorter duration [Fig.~\hyperref[fig:2]{2(b)}]. To reduce detection noise on the measured efficiency, we apply six consecutive Floquet pulses and infer the average efficiency per pulse. To account for shot-to-shot variation in pulse intensity, we extract the peak efficiency from a histogram of 100 shots by fitting to a model that includes pulse area noise and additive detection noise~\cite{Supplemental}. The observed peak efficiency is $\geq\!99.4\%$ for all detunings, compared to $99.63(4)\%$ for a resonant conventional $\pi$ pulse. The variation of the pulse efficiency as a function of detuning is explained by differences in spontaneous emission loss due to the corresponding pulse duration, indicating negligible losses from Doppler detuning. Our pulse efficiencies are comparable to recent atom interferometry results~\cite{Saywell2020_Raman}, as well as experiments with single atoms in optical tweezers~\cite{Madjarov2020}.

To demonstrate the advantage of Floquet atom optics over conventional pulses, we construct LMT Mach-Zehnder interferometers with momentum separation $N\,\hbar k$, where $N$ is the LMT order. The sequence consists of a $\pi/2$ beam splitter pulse, followed by a series of $\pi$ pulses from alternating directions, each of which transfers a net $2\,\hbar k$, and then a final $\pi/2$ pulse, as described in~\cite{Rudolph2020}. Note that there is no added interrogation time between the beam splitter and mirror pulses. For an LMT interferometer using Floquet atom optics, each pulse must be individually optimized for the instantaneous velocities of the interferometer arms. The pulse parameters $\delta$ and $t_\pi$ are determined by the analytic expressions above based on the Doppler detuning.  Since conventional pulses are marginally more efficient at the lowest velocities, we use conventional pulses until the arm velocities reach $26\,\hbar k$ and then switch to Floquet pulses for the majority of the sequence.

\begin{figure}
         \includegraphics[width=\linewidth]{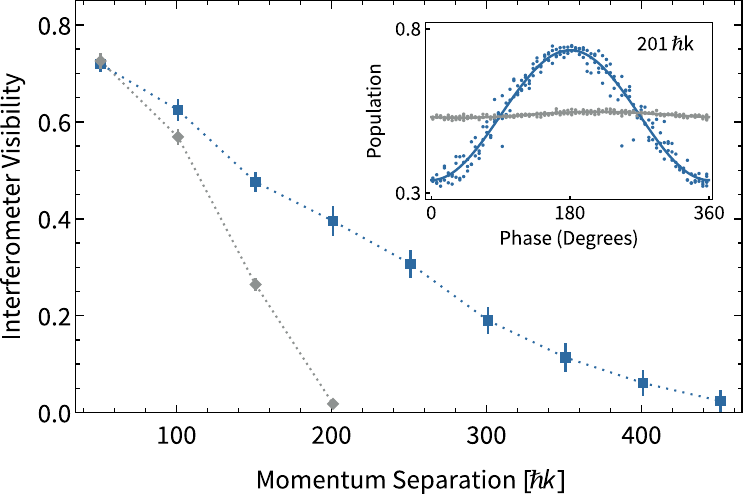}
     \caption{LMT interferometer visibility as a function of the momentum separation of the arms using Floquet atom optics (blue squares) and conventional $\pi$ pulses (gray diamonds). The dotted lines are a guide to the eye. The error bars are the standard deviation of the residuals of the sinusoidal fits. The inset is an example for a 201~$\hbar k$ interferometer, showing the normalized excited state population versus the interferometer phase for Floquet atom optics (blue) and conventional pulses (gray), as well as the associated fits (solid curves).  
     }
    \label{fig:3}
\end{figure}

The visibility of the interferometer is measured by scanning the phase of the final $\pi/2$ pulse and fitting the amplitude of the resulting sinusoidal interference fringe [see Fig.~\ref{fig:3} (inset)]. Each point in the scan is the population in a bin around the center of each interferometer output port, with bin size chosen to maximize the interferometer signal-to-noise ratio. When using conventional pulses, the visibility is lost for momentum separation above $200\,\hbar k$. In comparison, with Floquet atom optics we maintain interferometer visibility beyond $400\,\hbar k$ [see Fig.~\ref{fig:3}]. This is the highest momentum separation with sequential-pulse atom optics reported to date, even rivaling coherent acceleration with optical lattice-based methods~\cite{Gebbe2021, Pagel2020}. Additionally, we achieve state-of-the-art efficiency per $\hbar k$, competitive with other atom-optics techniques~\cite{Kovachy2012, Jaffe2018, McAlpine2020}.

It is illustrative to compare Floquet atom optics to other pulse engineering methods. In contrast to Floquet pulses, techniques such as composite pulses, ARP, and optimal control typically require at least twice the conventional $\pi$-pulse duration $t_0$~\cite{Levitt1986, Vitanov2001, Saywell2018}, leading to increased spontaneous emission loss.  Additionally, Floquet pulses can be simple to implement (e.g., AM using a mixer) without necessarily requiring an AFG. In the case of ARP, the state vector precesses multiple times as it adiabatically follows the torque vector, leading to unwanted dynamic phase accumulation~\cite{Kotru2015}. Floquet pulses instead result from a diabatic projection into the Floquet modes, avoiding systematic errors associated with extensive adiabatic precession. As mentioned, square-wave Floquet pulses may be considered a type of composite pulse sequence in which the phase toggles between 0 and $\pi$~\cite{Shaka1987}. In contrast to composite pulses that incorporate an unrestricted set of phases, Floquet atom optics are instantaneously mirror symmetric with respect to detuning. In the context of NMR, similar schemes have been considered to compensate for a distribution of pulse inhomogeneities, including Floquet modulation~\cite{Zax1988, Abramovich1993} and symmetric composite pulse sequences~\cite{Shaka1987}. Our approach differs in that we take advantage of strong coupling dynamics to tune the modulation offset $\delta$ and pulse time $t_{\pi}$ to find the exact $\pi$-pulse condition for a given detuning.  Finally, quantum optimal control methods~\cite{Werschnik_2007, Nielsen2010, Saywell2020} typically use many degrees of freedom to construct optimized phase and amplitude profiles, while Floquet $\pi$ pulses use only 3 degrees of freedom ($\Omega$, $\delta$, and $t_\pi$).

The performance of Floquet atom optics in LMT interferometers can be further improved by increasing the laser intensity. Operating at a Rabi frequency of $15~\text{MHz}$ would raise the theoretical limit for pulse efficiency to $99.9\%$, set by spontaneous emission loss due to the finite pulse duration. Floquet pulses approaching such efficiency can potentially support a momentum separation of $1000~\hbar k$. For a practical sensor, the interrogation time of the interferometer can be increased without adding spontaneous emission loss by shelving the atoms in the ground state~\cite{Rudolph2020}. Floquet atom optics are applicable to other systems where a discrete set of detunings must be addressed, such as dual-isotope LMT atom interferometers where the differential recoil velocity between the isotopes leads to Doppler detuning loss~\cite{Asenbaum2020, MAGIS}. Finally, these results have broad application in two-level systems exhibiting strong drive Rabi dynamics over a wide range of operating parameters.

This work was supported in part by NSF QLCI Award No. OMA-2016244, the Gordon and Betty Moore Foundation Grant GBMF7945, and the U.S. Department of Energy, Office of Science, QuantiSED Intitiative.

\end{document}